\documentclass[nofootinbib,twocolumn,preprintnumbers,floatfix]{revtex4-1}
\pdfoutput=1
\usepackage{amsmath,amsthm,amssymb,multirow,psfrag}
\usepackage{epsfig}
\usepackage{color}
\usepackage{soul}
\usepackage{appendix}

\usepackage{verbatim}
\usepackage{epsfig}
\usepackage{color}
\usepackage{comment}
\usepackage{slashed}
\usepackage{float}
\usepackage{bm}
\usepackage{commath}
\usepackage{multirow}
\usepackage{physics}
\usepackage{mathtools}
\usepackage{enumitem}

\usepackage{subcaption} 

\usepackage{graphicx}
\graphicspath{{Plots/}{FiguresV2/}{FiguresV3/}}

\usepackage[compat=1.1.0]{tikz-feynman} 
\usepackage{tikz} 
\usetikzlibrary{shapes,arrows,positioning,automata,backgrounds,calc,er,patterns}
\usepackage{feynmf}
\usepackage{tikz-network}
\usepackage{soul}
\usepackage[normalem]{ulem}
\setstcolor{red}
\tikzfeynmanset{warn luatex=false}

\begin{document}


\title{Freeze-in at stronger coupling }

\author{Catarina Cosme}
\email{ccosme@uc.pt}
 \affiliation{Univ Coimbra, Faculdade de Ci\^encias e Tecnologia da Universidade de Coimbra \\
and CFisUC, Rua Larga, 3004-516 Coimbra, Portugal}

\author{Francesco Costa}
\email{francesco.costa@uni-goettingen.de}
 \affiliation{Institute for Theoretical Physics, Georg-August University G\"ottingen, \\
  Friedrich-Hund-Platz 1, G\"ottingen D-37077, Germany}

\author{Oleg Lebedev}
\email{oleg.lebedev@helsinki.fi}
 \affiliation{Department of Physics and Helsinki Institute of Physics,\\
  Gustaf H\"allstr\"omin katu 2a, FI-00014 Helsinki, Finland \\}

\date{\today}

\begin{abstract}
The predictivity of many non-thermal dark matter (DM) models is marred by the gravitational production background. 
Nevertheless, if the reheating temperature $T_R$ is low, the gravitationally produced relics can be  diluted.
We study the freeze-in  dark matter production mechanism at temperatures below the dark matter mass.
In this case, the coupling to the thermal bath
has to be significant to account for the observed dark matter  relic density.  As a result, the direct DM detection experiments already probe such 
freeze-in models, excluding significant parts of parameter space.   
\end{abstract}

\maketitle


\section{\label{sec:Intro} Introduction}

The nature of dark matter is one of the most puzzling questions in modern physics. While the Weakly Interacting Massive Particle (WIMP) paradigm may potentially  explain the magnitude of the DM relic density,
it has come under pressure from ever-improving direct DM detection experiments \cite{LZ:2022ufs}. Although certain classes of WIMP models remain viable \cite{Gross:2017dan}, one is motivated to explore alternatives. An interesting possibility
is known as the ``freeze-in'' mechanism \cite{McDonald:2001vt,Kusenko:2006rh,Hall:2009bx}, in which case dark matter abundance builds up gradually due to its very weak coupling to the Standard Model (SM) thermal bath,
while DM itself remains non-thermal.
The main challenges for this 
framework are: (1) it is very difficult to test due to the small couplings, and (2) it assumes zero initial DM abundance, which is a strong assumption in a world with gravity. Concerning the second point, 
particles are efficiently produced by classical and quantum gravitational interactions during and immediately after  inflation, when the relevant energy scales are   not far from the Planck scale. 
This creates an important  background for non-thermal dark matter studies \cite{Lebedev:2022cic}.  
  
In our current work, we suggest how both of these problems could be addressed in a straightforward manner, at least in part of the parameter space.  The reheating temperature of the Universe, $T_R$, can be quite low, which dilutes the gravitationally produced relics. In this case, the dark matter mass may easily be above $T_R$ such that DM production is Boltzmann-suppressed and necessitates larger couplings, up to ${\cal O}(1)$, without thermalizing the dark sector. This opens up the possibility of direct detection of freeze-in dark matter, even in the simplest setups.

\section{\label{sec:motivation} Motivation: gravitational production background}

Typical models of non-thermal dark matter suffer from the problem of gravitational particle production background \cite{Lebedev:2022cic}. Feebly coupled particles are abundantly  produced by gravity during and immediately after inflation, thereby affecting their eventual density.  
To give an example, consider 
a feebly interacting  scalar $s$ of mass $m_s$ minimally coupled to gravity in an expanding Universe (see, e.g.\,\cite{Ford:2021syk}). 
As long as its mass is below the Hubble rate $H$, the field experiences quantum fluctuations.  Depending on the size of the self-coupling $\lambda_s$, these fluctuations reach an equilibrium value
according to the Starobinsky-Yokoyama distribution  \cite{Starobinsky:1994bd}. At the end of inflation, they can be interpreted as a condensate of the long wavelength modes, 
which can subsequently be converted into the particle  number density using simple scaling arguments \cite{Peebles:1999fz, Markkanen:2018gcw}.
Since the interactions of the scalar are feeble, it does not thermalize and the total particle number is conserved
at late times. When it becomes non-relativistic, $s$ can contribute significantly to the Universe's energy density.
 
The resulting abundance of $s$-relics is typically  very large for standard large-field inflation models \cite{Lebedev:2022cic}. Since the relics are relativistic at the end of inflation, their energy density can 
be ``diluted'' if the Universe expands as non-relativistic matter for a long enough period. This is the case, for example, when the energy density is dominated by the inflaton $\phi$ oscillating in a quadratic potential
${1\over 2 }m_\phi^2 \phi^2$ and having very small couplings to other fields. 
Since the inflaton decays at late times, e.g., via its small Higgs coupling $\phi H^\dagger H$,  the resulting   
reheating temperature is low, possibly in the GeV range.  
   
The consequent constraint on cosmological models is best formulated in terms of the relic abundance of the $s$-particles,
\begin{equation}
Y = {n \over s_{\rm SM}} ~~,~~  s_{\rm SM} = {2\pi^2 g_*\over 45} \, T^3   \;,
\end{equation}
where $g_*$ is the number of degrees of freedom in the SM and $n$ is the number density of the $s$-quanta. The observational constraint on the dark matter abundance is 
$Y_{\rm obs}= 4.4 \times 10^{-10} \; \left(   {{\rm GeV}\over m_s}  \right)$ \cite{Planck:2015fie}.
Requiring the $s$-relic abundance not to exceed that of dark matter,
the constraint on the duration of the non-relativistic expansion period 
reads \cite{Lebedev:2022cic}\footnote{This assumes that the scalar field reaches the Starobinsky-Yokoyama distribution \cite{Starobinsky:1994bd} during the de Sitter phase. Relaxing this assumption makes the bound somewhat weaker, without affecting any of the conclusions \cite{Lebedev:2022cic}.}
   \begin{equation}
          \Delta_{\rm NR} \gtrsim    10^7 \,    \lambda_s^{-3/4}         \left( {H_{\rm end}    \over M_{\rm Pl}     }\right)^{3/2}    \,     \left(        {m_s \over {\rm GeV}}       \right)       \, 
    \;,
  \label{m-lambda-NR}
\end{equation}
where $\lambda_s$ is the self-coupling ($\Delta V = {1\over 4 } \lambda_s s^4$), $H_{\rm end}$ is the Hubble rate at the end of inflation and the ``dilution'' factor
$\Delta_{\rm NR}$ characterizes the duration of the non--relativistic expansion period,
\begin{equation}
 \Delta_{\rm NR} \equiv
\left( {H_{\rm end}\over H_{\rm reh}}\right)^{1/2}  \simeq  {T_{\rm inst} \over T_R}\;.
 \end{equation}
Here the Hubble rate scales as $a^{-3/2}$  after inflation until reheating occurs, at which point it is given by  $H_{\rm reh}$.
The dilution factor can  also be written as the ratio of the reheating temperature in case of instant reheating, $T_{\rm inst} $,  and the actual reheating temperature, $T_R$.
Clearly, this factor can be very large, reaching $10^{18}$ in the extreme case of a 4 MeV reheating temperature \cite{Hannestad:2004px}.
To illustrate the strength of this constraint, let us take $H_{\rm end} \sim 10^{14}$ GeV characteristic of large field inflation models. In this case, the scalar masses only far below a GeV are allowed
for $\Delta_{\rm NR} \sim 1$. Therefore, a long enough non-relativistic expansion period is required to allow for stable relics with masses above 1 GeV.
    
Gravitational particle production continues during the inflaton oscillation epoch, albeit via a different mechanism. Inflaton oscillations induce an oscillating component in the scale factor $a(t)$, which results
in particle production \cite{Ema:2015dka}.  More importantly, quantum gravitational effects are expected to induce all gauge invariant operators
including Planck-suppressed couplings between the inflaton  $\phi $ and dark matter \cite{Lebedev:2022ljz}, e.g.
\begin{equation}
C \; {\phi^4 s^2 \over M_{\rm Pl}^2} \;,
\end{equation}  
where $C$ is a Wilson coefficient.
After inflation, $\phi$ undergoes coherent oscillations, which leads to efficient particle production. At weak coupling, the production is non-resonant and can be computed perturbatively \cite{Dolgov:1989us,Traschen:1990sw,Ichikawa:2008ne}, which yields
the constraint \cite{Lebedev:2022cic}
\begin{equation}
 \Delta_{\rm NR} \gtrsim 10^6\; C^2  \;{       \phi_0^8  \over    H_{\rm end}^{5/2}  \, M_{\rm Pl}^{11/2}} \; {m_s \over {\rm GeV}} \;,
  \label{c4-bound}
 \end{equation}
where $\phi_0$ is the inflaton field value at the end of inflation. Taking $\phi_0 \sim M_{\rm Pl}$ and $H_{\rm end} \sim 10^{14}$ GeV, we get a very strong bound
 $
 \Delta_{\rm NR} \gtrsim 10^{17 }\; C^2 \;  {m_s \over {\rm GeV}} \;.
 $
Unless the quantum gravity effects are well under control and $C \ll 1$,  $ \Delta_{\rm NR}$ is required to be extremely large for the GeV scale DM masses \footnote{Depending on the inflaton mass and $C$,
the $\phi^4 s^2 $ interaction can also lead to resonant DM production and important collective effects, which have been studied with the help of lattice simulations in \cite{Lebedev:2022ljz}.}. 
The corresponding reheating temperature would have to be very low, down to the GeV  range.  
Such a low-scale reheating can be achieved with a very small inflaton-Higgs coupling $\sigma_{\phi h} \phi H^\dagger H$, which yields $T_R \sim \sigma_{\phi h} \, \sqrt{M_{\rm Pl} / m_\phi}$.
 
These considerations apply more generally to dark matter with significant couplings, which however does not thermalize.  If the number density
of the  quanta produced  at high energy   is low, their interaction rate would be  below the Hubble scale. In this case, the system would never thermalize despite significant couplings. 

\section{\label{sec:motivation} Boltzmann-suppressed freeze-in} 

There is no observational evidence that the temperature of the SM thermal bath has ever been very high. In fact, the reheating temperature could be as low as 4 MeV \cite{Hannestad:2004px}.
 This motivates us to study the possibility that the dark matter mass is significantly above the reheating temperature.
 In this case, DM never thermalizes and can be produced via freeze-in. Since its production rate is Boltzmann-suppressed, the coupling to the SM thermal bath can be substantial, which opens up prospects for
 direct detection of frozen-in dark matter.  In what follows, we illustrate this statement with a simple model using the instant reheating approximation, which is adequate when
 the maximal and reheating temperatures are close (see Appendix for justification).
 
 Consider real scalar dark matter $s$ of mass $m_s $ which couples to the SM fields via the Higgs portal \cite{Silveira:1985rk,Patt:2006fw},
  \begin{equation}
 V(s) = {1\over 2}  \lambda_{hs }s^2 H^\dagger H + {1\over 2} m_s^2 s^2 \;.
 \end{equation}
 Suppose that the initial abundance of dark matter before reheating is negligible, while the reheating temperature is much lower than $m_s$.
 In this case, $s$ is produced  via scattering of energetic  SM quanta, although the process is Boltzmann suppressed. 
   
 The leading   production process is the Higgs  and vector boson pair annihilation into the scalars. For $m_s \gg m_h$,  these modes can be accounted for by 
 using four Higgs degrees of freedom in accordance with  
  the Goldstone  equivalence theorem.
 The DM number density $n$ is found via the Boltzmann equation,
  \begin{equation}
\dot n + 3H n =  \Gamma (h_i h_i \rightarrow ss) - \Gamma (ss \rightarrow h_ih_i)   \;,
\label{B}
  \end{equation}
where $\Gamma$ is the reaction rate per unit volume and $i=1,..,4$ labels the Higgs field components. At very weak coupling, the inverse reaction rate is negligible, however in general it can be significant.
  
If the temperature of the SM bath $T$ is far below $m_s$, only the particles at the Boltzmann tail, $E/T \gg 1$,    have enough energy for DM pair production. Therefore, the effects of quantum statistics 
\cite{Lebedev:2019ton,DeRomeri:2020wng}
can safely be neglected.
The    $h_i h_i \rightarrow ss$     reaction rate is given by \cite{Gondolo:1990dk}
\begin{eqnarray}
  \Gamma (h_i h_i \rightarrow ss)&=& \langle \sigma(h_i h_i \rightarrow ss)\,   v_r \rangle  \, n_h^2 \nonumber \\
  &=& {1\over (2\pi)^6} \;     \int \sigma v_r \, e^{-E_1/T} e^{-E_2/T}  d^3 p_1 d^3 p_2  \nonumber \\
  &=&  \frac{2\pi^{2}T}{\left(2\pi\right)^6}\int_{4m_{\rm s}^{2}}^{\infty}d\rm s\,\sigma\left(\rm s-4m_{h}^{2}\right)
  \nonumber\\
  &&
  \times\sqrt{\rm s}\,K_{1}\left(\frac{\sqrt{\rm s}}{T}\right)\;,
 \label{Gammahhss}
\end{eqnarray}
where $\sigma$ is the   $h_i h_i \rightarrow ss$ cross-section; $v_r$ is the relative velocity of the colliding quanta with energies $E_1,E_2$ and momenta $p_1,p_2$; $n_h$ is the Higgs boson number density;
$\langle ... \rangle $ denotes a thermal average;
${\rm s}$ is the Mandelstam variable, and
$K_1(x)$ is the modified Bessel function of the first kind.   
The cross-section is 
 \begin{equation}
\sigma = 4 \times {\lambda_{hs}^2 \over 32\,\pi\, {\rm s}  } \; {  \sqrt{{\rm s} - 4m_s^2}   \over   \sqrt{ {\rm s} - 4m_h^2} } \;,
\end{equation}
where we have included the Higgs multiplicity factor $4$ directly in $\sigma$.
At the threshold ${\rm s} = 4m_s^2$, the cross-section vanishes. Therefore, the integrand  peaks just above  ${\rm s} = 4m_s^2$ due to the Boltzmann suppression. At $z= \sqrt{s}/T \gg 1$, we may approximate 
$K_1 (z) \simeq \sqrt{\pi \over 2z } \; e^{-z} $. Then,  the fast varying functions in the integrand are $\sqrt{s-4m_s^2} $ and $e^{-\sqrt{s}/T}$, while the other factors can be approximated by constants, setting
${\rm s} = 4m_s^2$. The resulting integral reduces to the Gamma function $\Gamma(3/2)$ and for $m_s^2 \gg m_h^2 $ we get 
 \begin{equation}
  \Gamma   (h_i h_i \rightarrow ss)   \simeq  {\lambda_{hs}^2 T^3  m_s \over 2^7 \pi^4} \,   e^{-2m_s/T} \;,
    \end{equation}
for four Higgs degrees of freedom (d.o.f.). The reaction rate drops exponentially with temperature, as expected.
 
Let us assume for now that the coupling is weak enough and 
the inverse reaction $ss \rightarrow h_i h_i$ is unimportant. In this case, the Boltzmann equation can  easily be solved. Let us write it in the form
\begin{equation}
 {d\over dT} {n\over T^3} = - {\Gamma (h_i h_i \rightarrow ss)  \over HT^4} \;,
 \label{BoltzmannT}
\end{equation}
which assumes that the number of degrees of freedom in the SM stays approximately constant at relevant energies, so that   $a^3T^3 = $ const and     $dt = -dT /(HT)$.
Given that 
$H = \sqrt{ \pi^2 g_* \over 90}  {T^2 \over  M_{\rm Pl}}   $, the Boltzmann equation can be integrated analytically in terms of elementary functions. 
Starting with zero DM density at $T_R$, the eventual 
DM abundance    $Y= n/s_{\rm SM}$    is found to be
\begin{equation}
Y= {\sqrt{90} \, 45 \over  2^{9} \pi^7 \, g_*^{3/2}  }           \, {\lambda_{hs}^2 M_{\rm Pl} \over  T_R} \,e^{-2m_s/T_R}   \;,
\label{Y1}
\end{equation}
with 4 effective Higgs d.o.f. 
Imposing the   constraint $Y_{\rm obs}= 4.4 \times 10^{-10} \; \left(   {{\rm GeV}\over m_s}  \right) \;,$
we obtain 
 \begin{equation}
\lambda_{hs} \simeq 3 \times10^{-11}\;  e^{m_s/T_R} \, \sqrt{T_R \over m_s} \;,
\label{result0}
 \end{equation}
where we have taken $g_* \simeq 107$.
We observe that the required coupling is a function of the ratio $m_s/T_R$ and the smallness of the prefactor in (\ref{result0}) can easily be compensated by $e^{m_s/T_R}$.
Thus, order one couplings are allowed by this mechanism.  Clearly, the result also applies to high reheating temperatures as long as DM production is Boltzmann-suppressed.
In contrast, the coupling required  by the conventional   freeze-in production       is in the range of $10^{-11}$ \cite{Yaguna:2011qn}.
If $T_R \gtrsim m_s$, we recover this result.

An important ingredient in our analysis is the assumption of Higgs thermalization. 
It is certainly satisfied  at the GeV scale and above, mainly due to the reaction $\bar b b \rightarrow h$ whose
rate is above the Hubble scale, $\Gamma (\bar b b \rightarrow h ) > 3 H n_h$.
We also note that reactions of the type $\bar bb \rightarrow ss$ do not contribute significantly to DM production. The reason is that, although $b$-quarks are more abundant than the Higgses,
only particles at the Boltzmann tail $E/T \gg 1$ contribute (see Eq.\,\ref{Gammahhss}), exactly as in the Higgs case. Since there is an additional Yukawa coupling  and $(v_{\rm EW}/m_s)^2$ suppression compared to the Higgs reaction, such processes are unimportant.   

Finally, the $h_i h_i \rightarrow ss$ contribution at second order in $\lambda_{hs}$  is suppressed by the factor $\lambda_{hs} \, (v_{\rm EW}/m_s)^2 $, which makes it insignificant in the region of interest.
Analogously, the $s$-channel Higgs exchange contribution is suppressed by $\lambda_h v^2 / m_s^2$.  Our full numerical analysis shows that the above  approximation is adequate 
for $m_s^2 \gg m_h^2$.


\section{\label{sec:DM annih}Dark matter annihilation effect}

The $h_i h_i \rightarrow ss$ reaction is efficient only for a short period, after which its rate drops exponentially. During this period, it  populates the dark sector. Depending on the coupling,
the density of dark matter can be high enough to enable its annihilation in the SM states.  Let us study the evolution of DM abundance after the 
$h_i h_i \rightarrow ss$ mode has become inefficient.
The $s$-quanta are non-relativistic and 
their  annihilation reaction rate is given by
\begin{eqnarray} 
  \Gamma (ss \rightarrow h_ih_i)  =  \sigma (ss \rightarrow h_ih_i ) v_r \;n^2~~, ~~ 
  \nonumber \\
  \sigma (ss \rightarrow h_ih_i ) v_r=  4 \times {\lambda_{hs}^2 \over {64 \pi m_s^2} }
  \end{eqnarray}
for $m_s^2 \gg m_h^2$ and 4 Higgs d.o.f.
It is clear that if annihilation is efficient initially, $n$ drops quickly and the reaction stops. At late times, $ \Gamma (ss \rightarrow h_ih_i)   \propto n^2 \propto T^6$ is below $H n \propto T^5$,
and this reaction can be neglected in the Boltzmann equation.

Let us estimate the effect analytically. Eq.\,\ref{BoltzmannT} in our regime 
can be  written as 
\begin{equation}
 {d\over dT} {n\over T^3} =    \alpha \;\left(   {n\over T^3}     \right)^2    \;
 \label{ann-eq}
\end{equation}
where the constant $\alpha \equiv   \sqrt{ \pi^2 g_* \over 90}  \sigma (ss \rightarrow h_ih_i ) v_r / {  M_{\rm Pl}} $. 
Defining $T_* $ as the temperature at which DM production stops, we integrate this equation from $T_*$ to 0 and obtain the final abundance,
\begin{equation}
Y =      {  Y(T_*)    \over 1+   \beta\, Y(T_*) }          \;,
\label{YT*}
     \end{equation}  
where $\beta =  6.5 \times 10^{-3}g_*^{1/2} \, \lambda_{hs}^2 \, {M_{\rm Pl} T_* \over m_s^2}  $. If $\beta\, Y(T_*) \sim 1$, the annihilation effect is important. 
As $\lambda_{hs}$ grows, both $\beta $ and $Y(T_*)$ increase, leading to an upper bound on the DM abundance 
 \begin{equation}
 Y \lesssim {1\over \beta} \;
 \end{equation}
at large $\lambda_{hs}$.  Beyond some critical coupling, the abundance does not grow.
A rough estimate of this   coupling    can be obtained by approximating $Y(T_*)$ with the expression   (\ref{Y1}) and 
$T_* \sim T_R$, in which case  $\lambda_{hs } \sim 90 \times \sqrt{m_s \over M_{\rm Pl}} \,e^{m_s /(2 T_R)}$  for $g_* \simeq 107$. For example, at  
$T_R \sim 10 $ GeV and $m_s \sim 200$ GeV, the limiting coupling  is of order $10^{-1}$. Increasing the coupling beyond this value does not lead to a solution of $Y = Y_{\rm obs} $.
This is a feature of the Boltzmann equation of the type $y^\prime (x ) = -a\, e^{-b/x} / x^3 + c \, y^2$, with $a,b,c >0$ and $a\propto c$ as the coupling grows.
\begin{figure}[h!] 
\centering{
\includegraphics[scale=0.75]{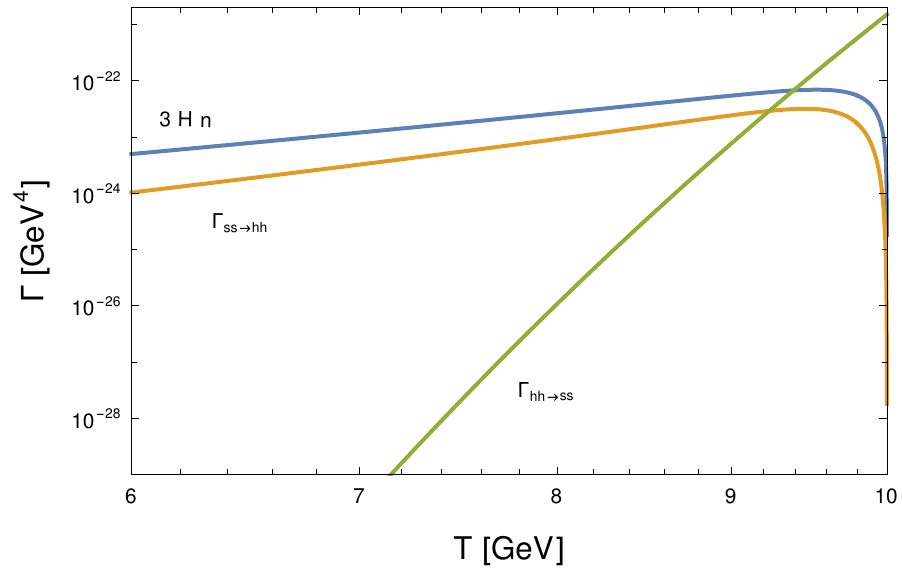}
\includegraphics[scale=0.75]{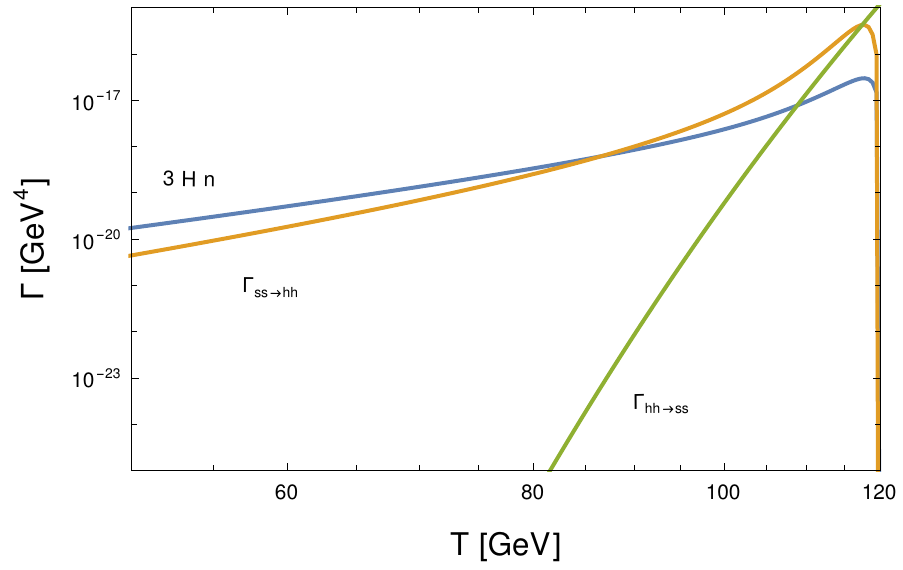}
}
\caption{ \label{RATES}
Evolution of the individual terms appearing in the Boltzmann equation. {\it Top:} $T_R=10$ GeV, $m_s=223.5$ GeV, $\lambda_{hs}= 0.05$.
{\it Bottom:} $T_R=120$ GeV, $m_s=2.81$ TeV, $\lambda_{hs}= 0.82$. 
}
\end{figure}
    
One may wonder whether significant 
$h_i h_i \rightarrow ss$ and $ss \rightarrow h_ih_i$ rates lead to thermalization of dark matter. 
However,  below the critical $\lambda_{hs}$,  this is not the case  
since these reactions are active at different times and 
\begin{equation}
  \Gamma (h_i h_i \rightarrow ss) \not=   \Gamma (ss \rightarrow h_ih_i) 
      \end{equation}  
apart from one point $T_*$. Below $T_*$, only the annihilation mode is active and $Y(T)$ evolves according to  (\ref{ann-eq}).
Thus, dark matter never equilibrates with the SM thermal bath. This can also be shown by comparing the numerical solution $Y(T)$ to the corresponding equilibrium value $Y_{\rm eq}(T)$.

This is illustrated in Fig.\,\ref{RATES}, which shows evolution of the reaction rates (per unit volume) and $3Hn$. The chosen parameters lead to the
correct DM relic abundance. We observe that the DM production mode is active only for a short period $\Delta T$ corresponding to a few percent of $T_R$ and its efficiency drops exponentially. In the top panel, the effect of DM annihilation is never significant,
whereas in the bottom panel, the annihilation mode is important for an extended period  and the resulting abundance is close to the maximal possible value at given $T_R$ and $m_s$. 
We note that in the temperature range where non-trivial dynamics take place,
the number of the SM degrees of freedom remains nearly constant  such that our approximation is adequate.  

In the vicinity of the critical coupling, 
the solution to $Y=Y_{\rm obs}$ develops a second branch with lower masses, which exhibits the scaling  
\begin{equation}
\lambda_{hs} \propto m_s \;.
\label{flat}
\end{equation}
On this branch,   $ \Gamma (h_i h_i \rightarrow ss) =   \Gamma (ss \rightarrow h_ih_i) $ for an extended period of time, and 
dark matter thermalizes. The relic abundance becomes independent of $T_R$ and is fully controlled by the annihilation cross-section, which leads to the above relation (see, e.g.\,\cite{Lebedev:2021xey}).

 
\section{\label{sec:Constraints} Constraints}

The main constraint on the model is due to the direct detection bounds, as is common to Higgs portal models (see a review in \cite{Lebedev:2021xey}). The DM scattering on nucleons is mediated by the Higgs, with the cross-section being  
\begin{equation}
\sigma_{sN} \simeq {\lambda_{hs}^2 f_N^2 \over 4\pi  } \, {m_N^4 \over  m_h^4 \,m_s^2} \;,
\end{equation}
where $f_N \simeq 0.3$ and $m_N\simeq 1\,$GeV.
The latest direct detection bound is due to LZ 2022    \cite{LZ:2022ufs}.
For instance, at $m_s =100$ GeV, the LZ bound is as strong as  $3\times 10^{-47}$cm$^2$, which constrains the coupling at the level of $10^{-2}$.  

The LHC and indirect detection bounds are superseded by the direct detection constraints in the region of interest \cite{Lebedev:2021xey}. Finally, DM is cold such that there are no significant structure formation constraints.
\begin{figure}[h!] 
\centering{
\includegraphics[scale=0.75]{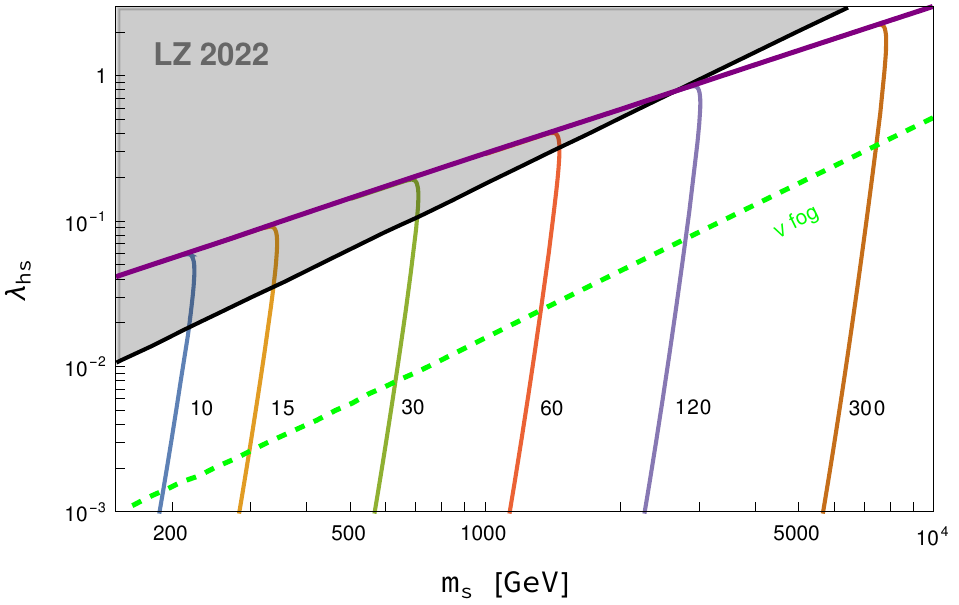}
}
\caption{ \label{par-space}
Parameter space of the Higgs portal freeze-in DM model. Along the curves, the correct DM relic abundance is reproduced. The curves are marked by the reheating temperature in GeV, while the purple line corresponds to thermal DM. The shaded area is excluded by
the direct DM detection experiment LZ 2022. The neutrino background for direct detection is represented by the dashed line ``$\nu$ fog''.}
\end{figure}
Fig.\,\ref{par-space} displays the most interesting region of parameter space of the model for electroweak scale reheating  temperatures ($T_R  \gtrsim 10$ GeV). Along the curves, the correct DM relic density is reproduced for a fixed $T_R$.
These results are obtained by solving numerically the full Boltzmann equation (\ref{B}). We observe that at lower couplings, the curves follow the pure freeze-in scaling law (\ref{result0}), while for larger couplings 
and lower masses, 
a second branch (\ref{flat})  develops due to DM thermalization. The latter corresponds to frozen-out dark matter.

The shaded area is excluded by the direct detection constraint. We see that LZ\,2022 is already sensitive to freeze-in DM masses up to 2-3 TeV.
Forthcoming experiments such as XENONnT \cite{XENON:2020kmp} and DARWIN \cite{DARWIN:2016hyl}
will be able to  probe much of the area above the ``$\nu$ fog'' line, which represents the neutrino background. 
Unlike in the standard WIMP case, the entire parameter space below the exclusion limit corresponds to viable DM models, and new experiments will continuously probe it.
The sensitivity to DM  can be extended beyond the neutrino floor using more sophisticated techniques, e.g. directional detection \cite{Ebadi:2022axg}. 
Clearly, the model allows for very small couplings and very large masses as well, yet such regions would be very difficult to probe, if possible at all.

Finally, let us note that certain freeze-in models with observable signatures have been built before, although these are more complicated and involve a  non-trivial dark sector with a specific spectrum
\cite{Hambye:2018dpi,An:2020tcg,Elor:2021swj,Bhattiprolu:2022sdd}.
The Higgs portal model at low $T_R$ was considered in
\cite{Bringmann:2021sth}, although it focussed on  light DM, which is strongly constrained by the Higgs invisible decay. 
In this work, we explore   a rather generic phenomenon of Boltzmann-suppressed  freeze-in dark matter, whose main signature would be  a direct detection signal in the upcoming experiments. 
This phenomenon was first observed in the context of spin 3/2
DM \cite{3/2}, where the reheating temperature is required to be below to avoid overproduction.

\vspace{1cm}


\section{\label{conclusion} Conclusion}

Gravitational particle production creates a background for many models of non-thermal dark matter, thereby impeding their predictivity.  This problem can be avoided 
if the  reheating temperature $T_R$ is relatively low. We study the simplest example of the Higgs portal dark matter, in which   DM is produced
via freeze-in at electroweak scale temperatures.  
If its  mass scale   lies above $T_R$, which we assume to be close to the maximal temperature, the production is Boltzmann suppressed and the coupling to the thermal bath can reach ${\cal O}(1)$, as required by the observed relic abundance. In this case, dark matter remains
non-thermal, while direct detection experiments provide a sensitive probe of such a  freeze-in scenario. In particular, LZ\,2022 already excludes a significant part of parameter space, while XENONnT and DARWIN
will probe lower couplings, down to the ``neutrino floor''. These conclusions also apply to more general freeze-in models that exhibit Boltzmann-suppressed DM production.


\begin{acknowledgments}
The work of F.C. is supported by the European Union's Horizon 2020 research and innovation programme under the Marie Sk{\l}odowska-Curie grant agreement No 860881-HIDDeN.
 C.C. is supported by the FCT project Grant No. IN1234 CEECINST/00099/2021, and partially by the FCT project Grant No. CERN/FIS-PAR/0027 /2021.
\end{acknowledgments}


\appendix*
\section{Evolution of the Standard Model bath temperature}

In this work, we have taken a low energy effective approach to dark matter creation, without specifying its possible UV completions.  
The crucial underlying assumptions  are that the temperature of Standard Model thermal bath has never been high and that dark matter production can be computed using instant reheating approximation.
The purpose of this appendix is to clarify under what circumstances this approach is justified.

The evolution of the Standard Model sector temperature depends on how exactly  the SM fields  are generated after inflation. Currently, this aspect of the early Universe dynamics remains essentially unconstrained and there are a number of viable options. For example, the SM sector can be generated directly via its interactions with the inflaton field $\phi$, e.g. by inflaton decay. An equally viable option is that 
the inflaton decays predominantly into other species, for instance, the right handed neutrinos $\nu_R$, whose decay  subsequently produces the   SM fields. 
Which option is realized in a particular model depends on the couplings strength of $\phi H^\dagger H$ versus that of $\phi \nu_R \nu_R$. The SM sector temperature evolution exhibits qualitatively different 
behaviour in the two cases: in the former case, it quickly reaches a high maximum and then decreases, while in the latter case, it reaches a plateaux and stays constant for a long time before decreasing.
It is the second option that motivates the freeze-in scenario proposed in our paper. 

Let us consider a general case of the SM radiation production via decay of some species $\chi$ with decay width $\Gamma_\chi$.  Denoting the SM energy density by $\rho$ and that of $\chi$ by $\rho_\chi$, we have
\begin{eqnarray}
&& \dot \rho + 4 H\rho= \Gamma_\chi \rho_\chi \;,  \nonumber \\
&& H= H_0 /a^m \;, \nonumber \\
&& \rho_\chi = \rho_\chi^0 / a^n \;. \label{system}
\end{eqnarray}
Here $n$ and $m$ parametrize the scaling of the Hubble rate and $\rho_\chi$, and are not necessarily related. The label ``0'' refers to the end of inflation, corresponding to $a=1$. The solution with the boundary condition 
$\rho (a=1)=0$ is 
 \begin{align}
 \rho(a) & =\frac{\Gamma_{\chi}\rho_{\chi}^{0}}{(4-n+m)H_{0}}\;\left[\frac{1}{a^{n-m}}-\frac{1}{a^{4}}\right]\nonumber \\
 & \rightarrow\frac{\Gamma_{\chi}\rho_{\chi}^{0}}{(4-n+m)H_{0}}\;\frac{1}{a^{n-m}}
  \label{rhoSM}
\end{align}
at $a\gg 1$ since $n-m <4$ for all cases of interest. 
Depending on $n-m$, 
  the SM energy density can grow, stay constant or decrease in time. So does the associated SM bath temperature,
  \begin{equation}
T\simeq       \left({30\over g_* \pi^2}\right)^{1/4}   \rho^{1/4} \;,
\end{equation} 
where $g_* $ is the number of the SM degrees of freedom. 

If $\chi$ corresponds to the inflaton, $n=2m$ and $\rho \propto a^{-m}$. In this case, the temperature decreases  and one expects the maximal temperature to be far above the reheating temperature, $T_{\rm max}\gg T_R$ (see e.g.  \cite{Co:2015pka,Calibbi:2021fld}). However, if $\chi$ is associated with feebly interacting right handed neutrinos, the situation is very different. The above calculation applied to the inflaton-$\nu_R$ system shows that
$\rho_\nu \propto 1/a^2$ for the $\phi^4$ inflaton potential and $\rho_\nu \propto 1/a^{3/2}$ for the $\phi^2$ case. Using $\nu_R$ as the source for the SM fields, e.g. via $\nu_R \rightarrow H \ell$, 
we therefore find 
\begin{equation} 
n-m=0 ~\Rightarrow~ T= {\rm const}
\end{equation}
as long as $\nu_R$ remains a subdominant energy component.  
Hence, one can have $T_{\rm max}$ and $T_R$ of similar size, and they can even coincide,
\begin{equation} 
 T_{\rm max} \simeq  T_R \;,
\end{equation}
\\
depending on the relation  between the inflaton and neutrino decay widths.  Therefore, the SM bath temperature has never been significantly higher than the reheating temperature,
justifying  the  scenario considered in this work.

The detailed analysis of the full $\phi-\nu_R-{\rm SM}$ system will be presented in our follow-up paper \cite{Cosme:2024ndc}. Below we list our most relevant results.

\begin{figure}[h!] 
{

\includegraphics[scale=0.59]{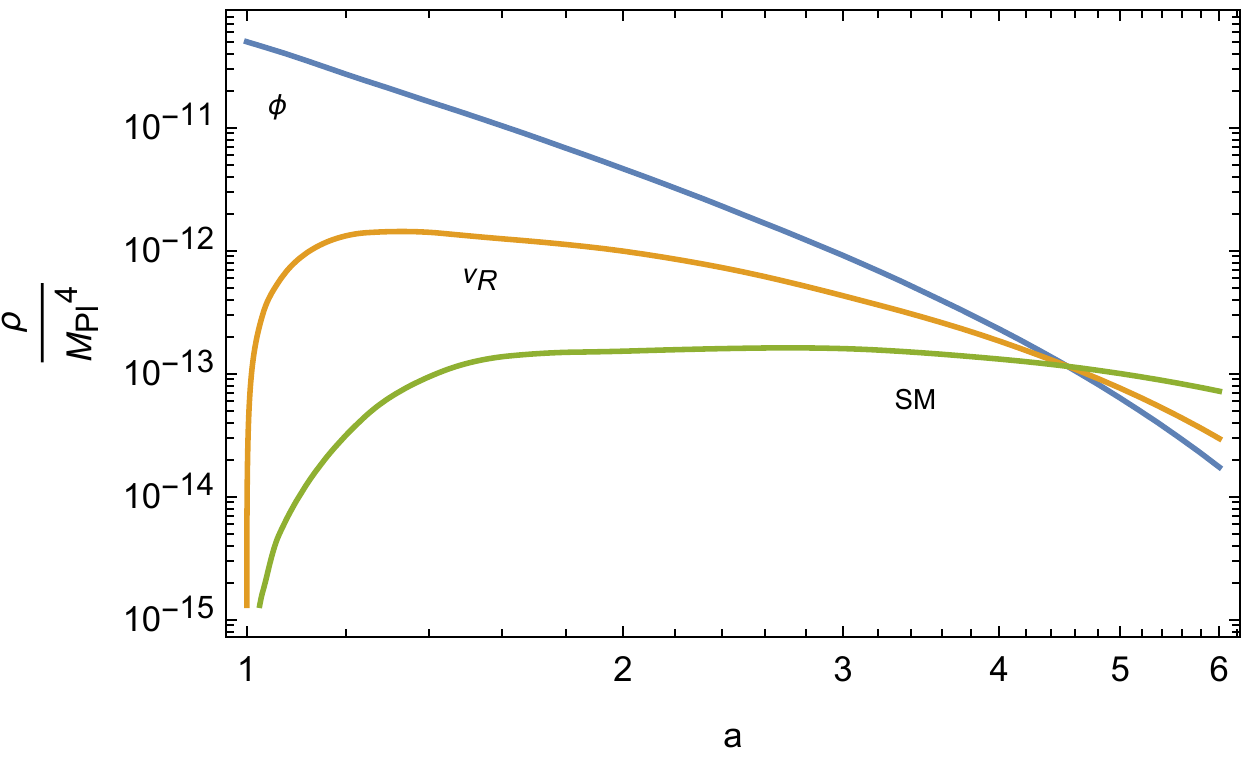}
}
\caption{ \label{appendix}
Evolution of the energy density components of the Universe for $\Gamma_\phi \sim \Gamma_\nu$. $\rho_\phi \propto a^{-3}$, $\rho_\nu \propto a^{-3/2}$, $\rho_{\rm SM} \propto a^0$. Reheating occurs at the point where the curves meet.}
\end{figure}

The numerical solution to the full system  of Boltzmann equations  for a quadratic inflaton potential is shown in Fig.\,\ref{appendix}. For illustration, we take $\Gamma_\phi \sim \Gamma_\nu$.
We observe that the SM energy density remains approximately constant until reheating, which realizes $T_{\rm max} \simeq  T_R $. 

In this case, dark matter is produced efficiently only in the vicinity of the reheating point. This is due to the dilution of the DM density produced earlier and Boltzmann suppression of its production later.
The same applies to the DM thermalization: it is most efficient at the reheating point due to the lower Hubble rate at the same SM temperature.
We find that the   results obtained using the complete system 
are very close to those presented in the paper. 

More generally, if $T_{\rm max}$ and $T_R $ are   different, the DM production is dominated by temperatures close to $T_{\rm max}$ 
and, to leading order, one can simply 
relabel
\begin{equation} 
 T_R \rightarrow T_{\rm max}  
\end{equation}
in the parameter space plots.  More precisely, the DM abundance receives a rescaling factor computed in \cite{Cosme:2024ndc}.

We therefore conclude that dark matter freeze-in at stronger coupling can be realized in a broad class of UV complete models, where the SM sector is produced by decay of a subdominant component in the energy density of the Universe.


\end{document}